\documentclass[a4paper,fleqn,usenatbib]{mnras}

\usepackage{newtxtext,newtxmath}

\usepackage[T1]{fontenc}
\usepackage{ae,aecompl}

\usepackage{times}
\usepackage{amssymb}
\usepackage{amsmath}
\usepackage{color}
\usepackage{bm}
\usepackage{graphicx}
\usepackage{tabu}
\usepackage{epsfig}
\usepackage{longtable,tabu}
\usepackage{url}
\usepackage{xcolor}
\usepackage{ulem}
\usepackage{supertabular}
\usepackage{caption}
\usepackage[multidot]{grffile}
\usepackage{subcaption}
\DeclareSymbolFont{matha}{OML}{txmi}{m}{it}
\DeclareMathSymbol{\varv}{\mathord}{matha}{118}

\def\mnras{MNRAS}
\def\aj{AJ}
\def\aap{A\&A}
\def\apj{ApJ}
\def\apjl{ApJ}
\def\apjs{ApJS}
\def\araa{ARA\&A}
\def\pasp{PASP}

\def\nat{Nature}

\def\lir{$L_{\rm IR}$}

\def\l1.4{$L_{\rm 1.4GHz}$} \def\s1.4{$S_{\rm 1.4GHz}$}

\def\gs{\mathrel{\raise0.35ex\hbox{$\scriptstyle >$}\kern-0.6em
\lower0.40ex\hbox{{$\scriptstyle \sim$}}}}
\def\ls{\mathrel{\raise0.35ex\hbox{$\scriptstyle <$}\kern-0.6em
\lower0.40ex\hbox{{$\scriptstyle \sim$}}}}
\def\m@th{\mathsurround=0pt }
\def\eqalign#1{\null\,\vcenter{\openup1\jot \m@th
 \ialign{\strut\hfil$\displaystyle{##}$&$\displaystyle{{}##}$\hfil
 \crcr#1\crcr}}\,}

\title[Giant star-forming clumps?]
{Giant star-forming clumps?}

\author[Ivison et al.]{R.\,J.~Ivison,\(^{\! 1}\) 
J.~Richard,\(^{\! 2}\)
A.\,D.~Biggs,\(^{\! 1}\)  
M.\,A.~Zwaan,\(^{\! 1}\)
E.~Falgarone,\(^{\! 3}\)
V.~Arumugam,\(^{\! 4,1}\)  
\and
P.\,P.~van~der~Werf\(^{\,5}\) 
and
W.~Rujopakarn\(^{6,7}\) 
\vspace*{1mm}\\
\(^1\) European Southern Observatory, Karl-Schwarzschild-Strasse~2, D-85748 Garching, Germany\\
\(^2\) Univ Lyon, Univ Lyon1, Ens de Lyon, CNRS, Centre de Recherche Astrophysique de Lyon UMR5574, F-69230, Saint-Genis-Laval, France\\
\(^3\) Laboratoire de Physique de l'ENS, ENS, Universit\'e PSL, CNRS,
Sorbonne Universit\'e, Universit\'e Paris-Diderot, Paris, France\\
\(^4\) Institut de Radioastronomie Millim\'etrique, 300 rue de la Piscine, F-38406 Saint-Martin d'H\`eres, France\\
\(^5\) Leiden Observatory, Leiden University, P.O.~Box 9513, 2300 RA Leiden, the Netherlands\\
\(^6\) Department of Physics, Faculty of Science, Chulalongkorn University, 254 Phayathai Road, Pathumwan, Bangkok 10330, Thailand\\
\(^7\) National Astronomical Research Institute of Thailand (Public Organisation), Don Kaeo, Mae Rim, Chiang Mai 50180, Thailand
}

\date{
Accepted 2020 March 12. Received 2020 March 12; in original form 2019
December 19
}

\pubyear{2020}

\begin{document}
\label{firstpage}
\pagerange{\pageref{firstpage}--\pageref{lastpage}}
\maketitle


\begin{abstract}
With the spatial resolution of the Atacama Large Millimetre
Array (ALMA), dusty galaxies in the distant Universe typically
appear as single, compact blobs of dust emission, with a median
half-light radius, $\approx1$\,kpc. Occasionally, strong
gravitational lensing by foreground galaxies or galaxy clusters
has probed spatial scales 1--2 orders of magnitude smaller,
often revealing late-stage mergers, sometimes with tantalising
hints of sub-structure.  One lensed galaxy in particular, the Cosmic
Eyelash at $z=2.3$, has been cited extensively as an
example of where the interstellar medium exhibits obvious,
pronounced clumps, on a spatial scale of $\approx100$\,pc.
Seven orders of magnitude more luminous than giant molecular
clouds in the local Universe, these features are presented as
circumstantial evidence that the blue clumps observed in many
$z\sim 2$--3 galaxies are important sites of ongoing star
formation, with significant masses of gas and stars. Here, we
present data from ALMA which reveal that the dust continuum of
the Cosmic Eyelash is in fact smooth and can be reproduced
using two S\'ersic profiles with effective radii, 1.2 and
4.4\,kpc, with no evidence of significant star-forming clumps
down to a spatial scale of $\approx80$\,pc and a star-formation
rate of $<3$\,M$_\odot$\,yr$^{-1}$.
\end{abstract}



\section{Introduction}
\label{introduction}

Interferometric submillimetre (submm) observations of distant, dusty,
star-forming galaxies (DSFGs, sometimes known as submm-selected
galaxies -- SMGs) --- intense starbursts with star-formation rates
(SFRs) in excess of 100\,M$_\odot$\,yr$^{-1}$ --- have revealed a
consistent morphological picture. Ignoring multiplicity and signatures
associated with galaxy interactions and mergers, of which there are
many examples, the thermal continuum emission from each is usually
dominated by a single, compact blob of dust -- expected to be largely
co-spatial with the molecular gas -- with a median half-light radius,
0.2--0.3\,arcsec or $\approx 1$\,kpc \citep[]{ikarashi15, ikarashi17,
  simpson15, hodge16, oteo16cplus, oteo17hires, rujopakarn16,
  gullberg18, rujopakarn19, ma19}.

In a handful of cases it has been possible to probe spatial scales
nearly an order of magnitude smaller, $\approx 150$\,pc or
$\approx 20$\,milliarcsec (mas), using the longest available
baselines, aided in one case by a bright, compact, in-beam calibrator
\citep{oteo17almacal2}. The findings are consistent -- compact blobs
of dust emission, occasionally multiple blobs suggestive of mid-stage
mergers \citep{iono16, tadaki18}. There have been glimpses of
sub-structure, interpreted by some as potential evidence for spiral
arms, bars and rings caused by tidal disturbances \citep{hodge19},
though some simulations and alternative analyses suggest we should be
cautious of their reality, or that they may instead be evidence of
mergers at a later stage \citep{hodge16, rujopakarn19}.

Strong gravitational lensing by foreground galaxies or galaxy clusters
allows us to probe spatial scales an order of magnitude smaller still,
at least in theory. The first of the three most celebrated cases is
that of H-ATLAS\,J090311.6$+$003906, or SDP.81, which lies at $z=3.0$
and is amplified by a single foreground galaxy ($\mu \approx 15$,
\citealt{dye15,rybak15a}, possibly with a $\sim 10^9$-M$_{\odot}$
dark-matter sub-halo -- \citealt{hezaveh16}). \citet{dye15} and
\citet{rybak15b} found evidence of a galaxy interaction \citep[as with
most bright SMGs ---][]{engel10}, in this case a late-stage merger
between a rotating disk of dusty gas and a neighbouring galaxy seen
only in the near-infrared (rest-frame optical).  Intriguingly, there
is evidence that the disk is fragmenting, since \citet{swinbank15}
identify up to five submm-emitting dust clumps, several of which can
be seen at multiple frequencies, on approximately the scale of the
synthesised beam ($\approx 150$\,mas, or $\approx 200$\,pc), which
supports more ambiguous evidence\footnote{Due to the possibility of
  excitation effects, or the patchy destruction of CO by cosmic rays
  \citep[e.g.][]{bisbas17}.}  from maps of CO(5--4) and CO(8--7) that
the gas distribution is clumpy.

Even more robust, though difficult to visualise because of the extreme
gravitational amplification that gives rise to its name, are the dozen
or more molecular clouds uncovered by recent 0.2-arcsec FWHM imaging
in CO(4--3) of the Cosmic Snake, at $z=1.036$, by
\citet{dz19}.  Each of these clouds is at least an order of magnitude
more massive and turbulent than those in the Milky~Way today, and
there is a substantial spatial disconnect between the gas and the
twenty clumps seen in {\it Hubble Space Telescope} imaging
\citep{cava18} which \citeauthor{dz19} had expected to detect in
CO.

Cited extensively as the definitive example of where the interstellar
medium (ISM) exhibits dusty star-forming clumps is the case of
SMM\,J21352$-$0102, or the Cosmic Eyelash, named due to its shape and its
proximity to the Cosmic Eye \citep{smail07,swinbank10}.  The Cosmic
Eyelash lies behind the $z = 0.325$ galaxy cluster, MACS\,J2135$-$01,
which amplifies it gravitationally by a factor, $\mu=37.5$.  With its
spectral energy distribution (SED) peaking at
$\lambda_{\rm obs} \approx 350\,\mu$m at a flux density,
$\approx 500$\,mJy \citep{ivison10fts} --- so typical intrinsically of
an SMG close to the confusion limit for a 10--15-m single dish --- and
with an SED that has proved invaluable for FIR/submm photometric
redshift estimation \citep[e.g.][]{gn19}, it was the first SMG
sufficiently bright to allow a blind redshift to be obtained,
$z=2.3$. This was determined by \citet{swinbank10} via detection of CO
$J=1$--0 using the Green Bank Telescope, a few months ahead of the
blind detection of CO $J=3$--2 and $J=5$--4 from SMM\,J14009+0252 by
\citet{weiss09}. The Cosmic Eyelash was also sufficiently bright to
allow FIR spectroscopy with the {\it Herschel} SPIRE FTS
\citep{george14, zhang18fts}. Early interferometric follow-up by
\citet{swinbank10}, using the eight-element Submillimeter Array (SMA)
in its most extended (VEX) configuration, provided evidence of at
least five and as many as eight bright, compact, dusty clumps.  The
most tempting lensing configuration suggested four on each side of a
caustic, each with an intrinsic spatial scale of $\approx 100$\,pc,
where the morphology of the molecular gas seen in later imaging by
\citet{swinbank11} was described as`broadly aligned' with the
continuum clumps.

\begin{figure}
\centering
\includegraphics[width=2.9in]{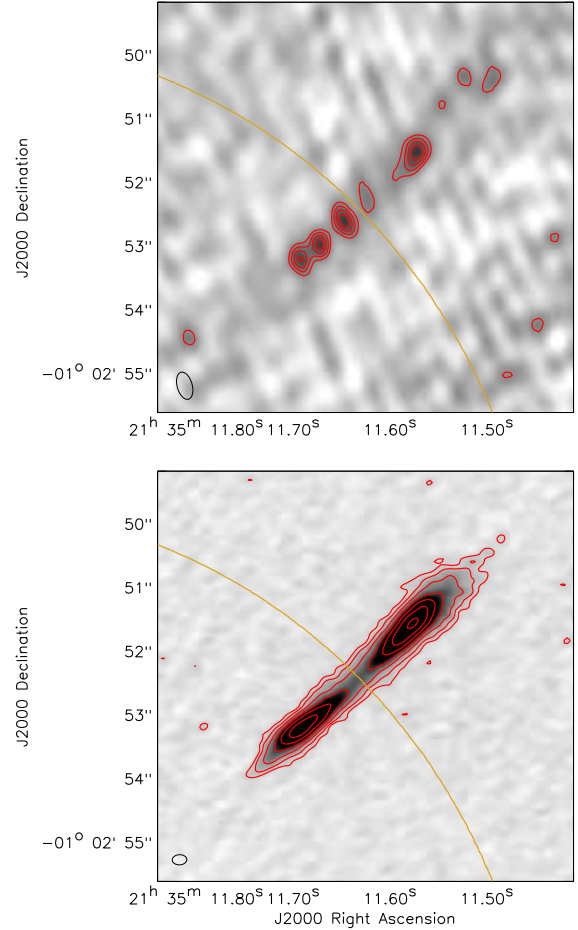}
\caption{{\it Top:} VEX-only SMA image of the Cosmic Eyelash from
  \citet{swinbank10}, where we have reproduced the contours shown in
  that paper as faithfully as possible, at 3, 4, 5 ... $\sigma$, where
  the r.m.s., $\sigma= 2.1$\,mJy\,beam$^{-1}$.  {\it Below:} our
  observed ALMA 251-GHz (band-6) continuum image (see
  \S\ref{sec:results}), which goes $\approx 40\times$ deeper than the
  SMA image after accounting for the shape of the SED, with contours
  at 3, 6, 12, 24 ... $\sigma$.  Each panel has the caustic
  illustrated, and the respective synthesised beam.}
\label{fig:sma_alma}
\end{figure}

\begin{figure*}
\centering
\includegraphics[width=6.3in]{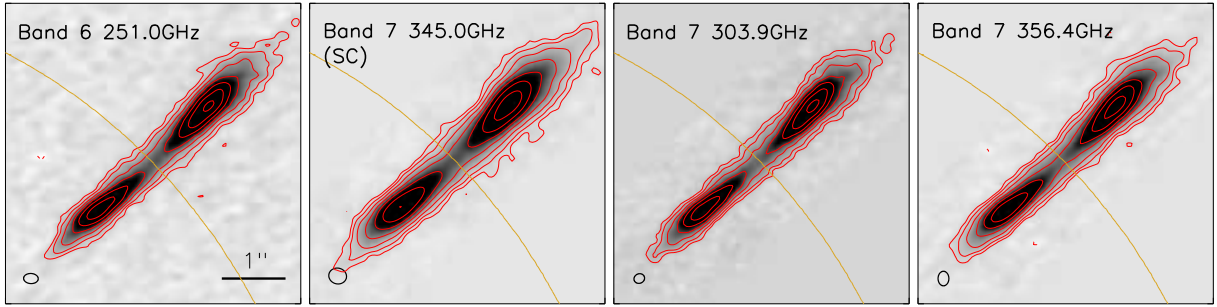}
\caption{ALMA images of the Cosmic Eyelash, as listed in
  Table~\ref{tab:obs} and described in \S\ref{sec:results}, displayed
  using linear greyscales from zero to the peak observed flux density,
  with contours at $-12$, $-6$, $-3$, 3, 6, 12 ... $\sigma$.  Each
  panel has the caustic illustrated, and also the synthesised beam,
  lower left. }
\label{fig:eyelashcont}
\end{figure*}

More than any other submm data, the discovery of these dust clumps in
the Cosmic Eyelash has been cited \citep[even quite recently
--][]{guo18, mg19} as circumstantial evidence that the giant
($\approx 0.1$--1\,kpc) off-centre clumps --- typically found in
broadband rest-frame ultraviolet (UV)--optical images of
$z\approx 1$--3 galaxies \citep[e.g.][]{cowie95, conselice04,
  elmegreen05, elmegreen08, elmegreen13, genzel11}, and especially in
IR-luminous galaxies \citep{calabro19} --- are important sites of star
formation, albeit perhaps short lived \citep[][cf.\
\citealt{bournaud14}]{genel12, kruijssen19, chevance19}.  Bright in
H$\alpha$ \citep{livermore12, livermore15}, these blue clumps are
thought to harbour significant star formation\footnote{One might ask why,
  since with adequate spatial resolution we generally find a
  disconnect between active star formation and blue light, and would
  anyway expect the dusty gas to be expelled rapidly, post-starburst,
  such that the ratio of UV to submm clumps may reflect the lifetimes
  for each phase.}  (though less than 10 per cent of the galaxy total
-- \citealt{guo18}) as well as significant quantities of stars
\citep[$M_\star \approx 10^7-10^9$\,M$_\odot$ -- ][cf.\
\citealt{wuyts12, cava18, zanella19, larson20}]{guo12, guo15, guo18,
  dz17, dz18} and any residual molecular gas from which those stars
formed.  On the other hand, simulations in UV and H$\alpha$ light
\citep[e.g.][]{tamburello17, mg19} and some data at longer wavelengths
--- less susceptible to the pernicious effects of dust --- suggest
that some of the best-known examples of star-forming clumps may have
masses and sizes that have been over-estimated and are likely rather
insignificant, plausibly even the result of patchy dust obscuration,
e.g.\ UDF11 in \citet{rujopakarn16} and UDF6462 in \citet{cibinel17}.

In this paper we present new observations obtained using the Atacama
Large Millimetre/Submillimetre Array (ALMA) which reveal that around
99 per cent of the dust continuum emission from the Cosmic Eyelash is
in fact distributed smoothly.  This paper is organised as follows:
\S\ref{sec:observations} describes the observations and data
reduction, and \S\ref{sec:results} presents the fundamental
result. In \S\ref{sec:lensmodel} we discuss our lens modelling of
the observed images.  We summarise and draw conclusions in
\S\ref{sec:summary}.  Throughout, we adopt the \citet{chabrier03}
initial mass function and a standard $\Lambda$-CDM cosmology with
$\Omega_{\rm m} = 0.3$, $\Omega_\Lambda = 0.7$ and
$H_0 = 70$\,km\,s$^{-1}$\,Mpc$^{-1}$, where 1\,arcsec at $z=2.3$
corresponds to 8.2\,kpc.

\section{Observations and data reduction}
\label{sec:observations}

The Cosmic Eyelash has been observed several times with ALMA,
predominantly in bands~6 and 7.  From these, we have selected a subset
with good sensitivity ($\le 50\,\mu$Jy\,beam$^{-1}$) and angular
resolution ($\Delta \theta \le 0.3$\,arcsec). Although mostly designed
to observe various molecular transitions, these data contain a
significant fraction of line-free channels that allow sensitive
continuum maps to be made.  One project used ALMA's maximum-bandwidth
($\Delta \nu = 7.5$\,GHz) `single-continuum' (SC) mode and --- although a
little less sensitive than the others, due to less observing time ---
contains both an extended and compact configuration and is thus
particularly sensitive to low-brightness extended emission.  All
observations were conducted in dual-polarisation mode with
low-spectral-resolution time-division mode (TDM) spectral windows,
i.e.\ with 2\,GHz of usable bandwidth. See Table~\ref{tab:obs} for a
summary of the band-6 and -7 ALMA observations considered for our
study.

Data reduction was carried out using the Common Astronomy Software
Application package, with calibration performed using the ALMA Science
Pipeline.  Contamination from molecular lines was identified by
combining all baselines to produce a spectrum, with affected channel
ranges then flagged.  Imaging was performed subsequently, using
\textsc{tclean} with a Briggs weighting scheme ({\sc robust} $=-0.5$).
Self-calibration was used (first in phase, then in amplitude and
phase) to produce the final continuum maps.  Each configuration of the
2012.1.01029.S data was mapped and self-calibrated separately.
Additional self-calibration of the combined datasets was necessary to
correct for small errors in the relative astrometry and flux-density
scales.

The most sensitive map was obtained from the band-6 data (rest-frame
360\,$\mu$m, where we probe emission from cold dust), published
previously as part of a survey of luminous, dusty galaxies in the
CH$^+$ line \citep{falgarone17}. The r.m.s.\ noise level was
$\sigma = 21\,\mu$Jy\,beam$^{-1}$ and the synthesised beam measured
$0.23 \times 0.16$\,arcsec$^2$ (FWHM), with the major axis at a
position angle (PA, measured East of North) of $93^{\circ}$.

In band~7, the map with the best sensitivity and highest angular
resolution was that produced from the 2012.1.00175.S data, intended
originally to trace OH$^+$ and H$_2$O (average frequency, 303.9\,GHz;
rest-frame 300\,$\mu$m).  A continuum map of these data has
already been published by \citet{indriolo18} but our map has a
significantly higher dynamic range and a sensitivity of
$\sigma = 27\,\mu$Jy\,beam$^{-1}$.  The synthesised beam was somewhat
smaller than that of the band-6 map, $0.17 \times 0.15$~arcsec$^2$ at
PA = $66^{\circ}$. A second dataset from the same project, targeting
H$_2$O at a higher frequency, 356.4\,GHz, produced a similar map,
though not quite as sensitive. The synthesised beam of the band-7
pure-continuum map was competitive with the other maps,
$0.28 \times 0.25$\,arcsec$^2$, PA = $104^{\circ}$, with a sensitivity
of $\sigma = 60\,\mu$Jy\,beam$^{-1}$.

\begin{table}
  \centering
  \caption{ALMA observations of the Cosmic Eyelash in bands~6 and 7.}
  \label{tab:obs}
  \begin{tabular}{lcccc} \\ \hline
ALMA & Detected & $\nu_0$ $^{\rm (b)}$& $\sigma$ /$\mu$Jy & Beam $^{\rm (d)}$\\
project& species $^{\rm (a)}$ & /GHz & beam$^{-1}$ $^{\rm (c)}$& /mas$^2$ \\ \hline
    2012.1.00175.S & OH$^+$, H$_2$O & 303.9 & 27 & $174 \times
                                                   148^{\rm (e)}$ \\
    2012.1.00175.S & H$_2$O         & 356.4 & 48 & $232 \times 167$ \\
    2012.1.01029.S & SC             & 345.0 & 60 & $282 \times 246$ \\
    2016.1.00282.S
     & CH$^+$         & 251.0 & 21 & $225 \times 160$ \\
    \hline
  \end{tabular}

  Notes: (a) SC refers to a `single-continuum' set up; (b) average
  frequency, after flagging of line-contaminated channels; (c)
  continuum sensitivity; (d) synthesised beam size, FWHM; (e) for the
  average magnifications across the Cosmic Eyelash, this corresponds to
  linear scales along the major and minor axes of 130 and
  820\,pc in the source plane, respectively.
\end{table}

\begin{figure*}
\centering
\includegraphics[width=6.3in]{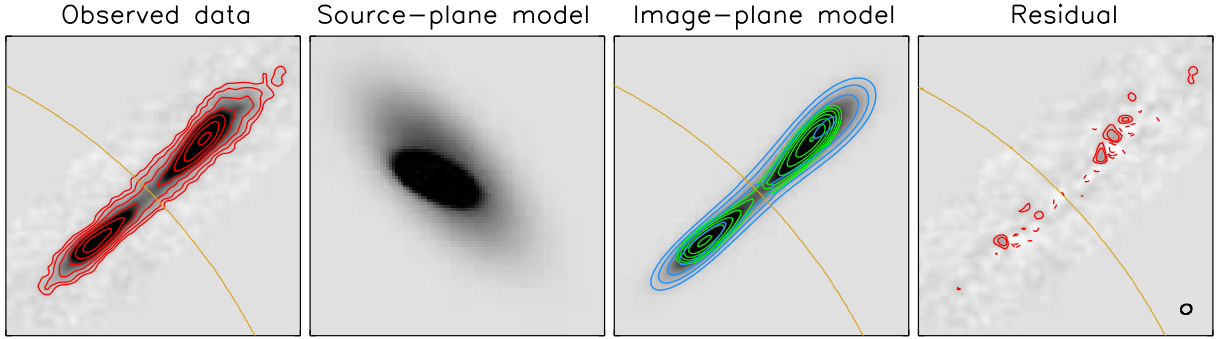}
\caption{{\it Left:} Our most sensitive ALMA image of the Cosmic
  Eyelash, with the finest angular resolution, at 303.9\,GHz, from the
  third panel of Fig.~\ref{fig:eyelashcont}. Contours are plotted at
  $-12$, $-6$, $-3$, 3, 6, 12 ... $\sigma$. {\it Middle left and
    right:} source- and image-plane models, respectively -- see
  \S\ref{sec:lensmodel} -- where the two best-fitting S\'ersic
  profiles are contoured separately (green and blue) in the image plane, and the
  source-plane panel is 11.5\,kpc across. {\it Right:} residuals,
  plotted with the same greyscale range and contours as the 
  observed data and image-plane model. The synthesised beam is shown,
  lower right.  Adopting the criteria of \citet{walter16} to assess
  the fidelity of the residual peaks, we find no reliable structure in
  the residual map.}
\label{fig:model}
\end{figure*}

\section{Results}
\label{sec:results}

We quickly and simply illustrate the purpose of this paper in
Fig.~\ref{fig:sma_alma}, which shows the SMA image\footnote{The SMA
  345-GHz (rest-frame 260\,$\mu$m) image of \citet{swinbank10} had an
  r.m.s.\ noise level of 2.1\,mJy\,beam$^{-1}$
  ($0.33\times 0.21$\,arcsec$^2$, PA = $15^\circ$).} of
\citet{swinbank10} alongside our deep ALMA band-6 continuum image.
The ALMA image is $\approx 40\times$ deeper than the SMA image, even
after accounting for the $2.6\times$ drop in observed dust emission
between 345 and 251\,GHz.  In all important respects the ALMA band-6
image has the same morphological characteristics as our band-7 imaging
(see Fig.~\ref{fig:eyelashcont}), which is more than $50\times$ deeper
than the SMA image, with a smaller and more symmetric synthesised
beam.  On the scales probed here, roughly 200\,mas in the image plane,
the dust continuum emission from the Cosmic Eyelash is remarkably
smooth, not clumpy.

Spatially resolved analysis performed at the positions of the clumps
identified by \citet{swinbank10}, which we have shown here to be
spurious, e.g.\ the work presented by \citet{swinbank11, danielson11,
  danielson13} and \citet{thomson15}, must be viewed in this context.
The clumpy structure presented by \citet{swinbank10} is believed to
have been generated by applying the {\sc clean} algorithm to noisy
long-baseline SMA data, amplifying features with low signal-to-noise
ratios (SNRs), where the remarkable symmetry of the resulting
structure about the likely caustic lent credibility to a clumpy
morphology that we show here to be spurious.  Similarities between the
molecular gas morphology presented by \citet{swinbank11} and the
spurious continuum clumps was the result of low SNR, as illustrated by
the simulations of \citet{hodge16}, which showed that high-resolution
low-SNR interferometric observations yield a clumpy distribution when
there are no clumps. Faced with such data, the lesson here is that an
analysis like that of \citeauthor{hodge16} should always be
undertaken, to gauge the reality of the clumps.

\section{Lens modeling}
\label{sec:lensmodel}

We have produced an updated version of the parametric mass model of
the MACS\,J2135$-$01 cluster core described in \citet{swinbank10}
using {\sc
  lenstool}\footnote{\url{https://projets.lam.fr/projects/lenstool/wiki}}
\citep{jullokneib09}. We take the centroids of the ALMA image pair in
Figs~\ref{fig:sma_alma}--\ref{fig:eyelashcont} as constraints.

We have used this lensing model to derive a parametric model of the
source morphology at the origin of the continuum emission.  We took a
forward-model approach, assuming a S\'ersic profile for the source,
convolving by the ALMA beam and re-gridding to the same pixel grid in
the image plane.  The source parameters (centroid, PA, axis ratio and
FWHM) were optimised while keeping the mass model fixed.  Because of a
small mis-match in the lens model,\footnote{With our parametric model
  for the cluster- and galaxy-scale mass components, the two images
  are reproduced with a small ($\sim 0.02$\,arcsec) offset, such that
  sending both images back to the source plane yields a small
  mis-match in position.} to reproduce both images simultaneously we
performed the fit on each image
independently, using the variations in the recovered parameters as an
estimate of systematic errors due to the lens model.

The best source parameters with a single S\'ersic profile reproduced
the observed configurations well, with significant ($>50\sigma$)
residuals near the core of each image, symmetrical about the critical
line.  Adopting a more complex parameterised source, as is becoming
routine with high-fidelity ALMA data \citep[e.g.][]{rujopakarn19} ---
this time comprising two independent S\'ersic profiles, as might be
expected for a merger-induced dusty starburst \citep{engel10} or for a
star-forming disk with a central starburst --- the resulting best-fit
source parameters gave two components lying very close in central
position (within 0.01\,arcsec or $\approx 80$\,pc in the source
plane), but having large differences in S\'ersic index and effective
radii.  The first one was rather extended ($R_{\rm e} \sim 4.4$\,kpc)
while the second one was brighter and more compact
($R_{\rm e}\sim 1.2$\,kpc), with a small S\'ersic index in both cases
\citep[$n\sim 0.5$, so at the low end of the range found by][but
consistent, as are the effective radii]{hodge16}. Table~\ref{tab:fits}
shows the best-fit parameters for each component.

Fig.~\ref{fig:model} presents our best ALMA continuum image of the
Cosmic Eyelash (Band~7, 303.9\,GHz) alongside the respective best-fit
source- and image-plane models, and residuals for the two-component
S\'ersic fit, where the observed map and the model are plotted with
the same linear scaling (from zero to the peak observed flux density)
as the residuals.

We found no mirrored sub-structure in the residual map and a brightest
peak of 260\,$\mu$Jy, roughly $10\sigma$ above the noise; the deepest
negative peaks reach 360\,$\mu$Jy which suggests --- along with the
lack of mirroring --- that the sub-structure we see is not real.  We
followed the approach of \citet{walter16} to assess the fidelity of
the residual peaks as a function of SNR, albeit needing to adopt large
SNR bins, searching for both positive and negative peaks.  We found no
reliable candidates, even at $10\sigma$: the fidelity of the brightest
peaks was never better than 50 per cent.  The residual peaks are all
approximately consistent with the size of the synthesised beam, i.e.\
they are unresolved down to $\approx 80$\,pc in the source
plane\footnote{Magnification varies spatially across the image plane,
  ranging from 3--$13\times$ along the major axis (mean,
  $\sim10\times$), and $\sim1.6\times$ along the minor axis.}  along
the major axis. If we scale the maximum positive residual, which is
magnified by roughly $8\times$ and $1.6\times$ along the major and
minor axes, to the well-sampled SED of the Cosmic Eyelash, we find
that its rest-frame 8--1000-$\mu$m luminosity cannot exceed
$24\times 10^9$\,L$_\odot$.  Adopting the traditional conversion from
\lir\ to SFR \citep[e.g.][]{ke12} --- noting that recent evidence for
a top-heavy stellar initial mass function in starbursts \citep[][cf.\
\citealt{romano19}]{zhang18, schneider18, motte18} would reduce these
SFR limits significantly --- then corresponds to a {\it maximum}
`clump SFR' of 2.6\,M$_\odot$\,yr$^{-1}$, around 1 per cent of the
total for the Cosmic Eyelash, at the low end of the range of SFRs
reported for clumps in star-forming galaxies at $z\sim 1$--3
\citep[e.g.][]{zanella19} and consistent with the values reported via
H$\alpha$ observations of strongly lensed galaxies at $1<z<4$
\citep{livermore12,livermore15}.  Adopting the extreme starburst SED
of Arp\,220, our limit moves $1.6\times$ higher.

\begin{table}
\centering
\caption{Best-fit parameters for the source model to the Band-7
  303.9-GHz image, with extended/compact components
  listed top/bottom, respectively.}
\begin{tabular}{ccccc}
\hline
$R_{\rm e}$ & Axis&PA&  Total flux   & S\'ersic$_{\rm n}$\\
/kpc & ratio& /deg &/mJy & \\
\hline
$4.42\pm 1.21$ & 0.46 & $134\pm4$ & $6.2\pm0.9$& $0.45\pm0.22$\\ 
$1.23\pm 0.02$ & 0.44 & $155\pm6$ & $8.3\pm1.3$& $0.51\pm0.06$\\
\hline
\end{tabular}
\label{tab:fits} 
\end{table}

\section{Summary}
\label{sec:summary}

We present sensitive, high-spatial-resolution ALMA continuum imaging
of the Cosmic Eyelash, at $z=2.3$, which has been cited extensively as
an example of where the interstellar medium exhibits obvious,
pronounced clumps, with spatial scales of $\approx 100$\,pc, and where
these clumps are cited regularly as circumstantial evidence that the
blue clumps observed in UV--optical images of many $z=2$--3 galaxies
are important sites of ongoing star formation, with significant masses
of stars and gas.

Our images reveal that the dust continuum emission from the Cosmic
Eyelash is smoothly distributed and can be reproduced using two
coincident S\'ersic profiles with effective radii, 1.2 and 4.4\,kpc,
with no evidence of significant star-forming clumps down to a spatial
scale of $\approx 80$\,pc, with rest-frame 8--1000-$\mu$m luminosities
below $24\times 10^9$\,L$_\odot$ and individual SFRs no higher than 1
per cent of the total, so $<2.6$\,M$_\odot$\,yr$^{-1}$.

\section*{Acknowledgements}

Sincere thanks to the anonymous referee whose suggestions improved
this paper significantly.
JR acknowledges support from the ERC Starting Grant
336736-CALENDS.
Funded by the Deutsche Forschungsgemeinschaft (DFG, German Research
Foundation) under Germany's Excellence Strategy -- EXC-2094 --
390783311.

ALMA is a partnership of ESO (representing its member states),
NSF (USA) and NINS (Japan), together with NRC (Canada), MOST and
ASIAA (Taiwan), and KASI (Republic of Korea), in cooperation with
the Republic of Chile. The Joint ALMA Observatory is operated by
ESO, AUI/NRAO and NAOJ. This paper relies on ALMA data from projects:\\
\noindent
{\small ADS/JAO.ALMA\#2012.1.00175.S, ADS/JAO.ALMA\#2012.1.01029.S,\\
ADS/JAO.ALMA\#2013.1.00164.S, ADS/JAO.ALMA\#2016.1.00282.S.}



\bibliographystyle{mnras}
\bibliography{rji} 


\label{lastpage}
\end{document}